\documentclass[10pt,aps,prd,superscriptaddress,nofootinbib,nobibnotes,longbibliography,floatfix,twocolumn]{revtex4-2}

\usepackage{bm}
\usepackage{mathtools,
amsmath,
amssymb,
amsfonts,
mathrsfs,
chngcntr,
multirow}

\let\cc\corresponds
\let\corresponds\relax
\usepackage{mathabx}
\let\corresponds\cc

\usepackage[utf8]{inputenc}
\usepackage[T1]{fontenc}
\usepackage[dvipsnames]{xcolor}
\usepackage[unicode]{hyperref}
\hypersetup{colorlinks=true, citecolor=MidnightBlue,
            linkcolor=MidnightBlue, urlcolor=MidnightBlue, linktocpage=true}
\usepackage[normalem]{ulem}
\usepackage{orcidlink}
\usepackage[capitalize]{cleveref}
\usepackage{float}
\usepackage{comment}

\usepackage{orcidlink}

\newcommand{\orcid}[1]{\href{https://orcid.org/#1}{\includegraphics[width=10pt]{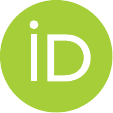}}}

\newcommand{\qm}[1]{``#1''}
\newcommand{\dd}{{\rm d}}

\begin{document}

\title{Confronting eikonal and post-Kerr methods with numerical evolution of \\ scalar field perturbations in spacetimes beyond Kerr}

\author{Ciro De Simone \orcid{0009-0004-0610-1686}}
\email{ciro.desimone@unina.it}
\affiliation{Dipartimento di Fisica \qm{E. Pancini}, Università di Napoli \qm{Federico II}, Complesso Universitario di Monte Sant' Angelo, Via Cinthia Edificio 6, I-80126 Napoli, Italy}
\affiliation{Istituto Nazionale di Fisica Nucleare, Sezione di Napoli, Complesso Universitario di Monte Sant' Angelo, Via Cinthia Edificio 6, 80126 Napoli, Italy}
\affiliation{Theoretical Astrophysics, IAAT, University of T\"ubingen, D-72076 T\"ubingen, Germany}

\author{Sebastian H.\,V\"olkel \orcid{0000-0002-9432-7690}}
\email{sebastian.voelkel@aei.mpg.de}
\affiliation{Max Planck Institute for Gravitational Physics (Albert Einstein Institute), \\ Am M\"uhlenberg 1, 
D-14476 Potsdam, Germany}

\author{Kostas D.\,Kokkotas \orcid{0000-0001-6048-2919}}
\email{kostas.kokkotas@uni-tuebingen.de}
\affiliation{Theoretical Astrophysics, IAAT, University of T\"ubingen, D-72076 T\"ubingen, Germany}

\author{Vittorio De Falco \orcid{0000-0002-4728-1650}}
\email{vittorio.defalco-ssm@unina.it}
\affiliation{Ministero dell'Istruzione e del Merito (M.I.M., ex M.I.U.R.), Italy}

\author{Salvatore Capozziello \orcid{0000-0003-4886-2024}}
\email{capozziello@na.infn.it}
\affiliation{Dipartimento di Fisica \qm{E. Pancini}, Università di Napoli \qm{Federico II}, Complesso Universitario di Monte Sant' Angelo, Via Cinthia Edificio 6, I-80126 Napoli, Italy}
\affiliation{Istituto Nazionale di Fisica Nucleare, Sezione di Napoli, Complesso Universitario di Monte Sant' Angelo, Via Cinthia Edificio 6, 80126 Napoli, Italy}
\affiliation{Scuola Superiore Meridionale,  Via Mezzocannone 4, I-80134 Napoli, Italy}

\begin{abstract}
The accurate computation of quasinormal modes from rotating black holes beyond general relativity is crucial for testing fundamental physics with gravitational waves. 
In this study, we assess the accuracy of the eikonal and post-Kerr approximations in predicting the quasinormal mode spectrum of a scalar field on a deformed Kerr spacetime. 
To obtain benchmark results and to analyze the ringdown dynamics from generic perturbations, we further employ a 2+1-dimensional numerical time-evolution framework. 
This approach enables a systematic quantification of theoretical uncertainties across multiple angular harmonics, a broad range of spin parameters, and progressively stronger deviations from the Kerr geometry. 
We then confront these modeling errors with simple projections of statistical uncertainties in quasinormal mode frequencies as a function of the signal-to-noise ratio, thereby exploring the domain of validity of approximate methods for prospective high-precision black-hole spectroscopy. 
We also report that near-horizon deformations can affect prograde and retrograde modes differently and provide a geometrical explanation. 
\end{abstract}

\maketitle

\section{Introduction}

One decade after the first direct detection of gravitational waves (GWs) from the binary black hole (BH) merger GW150914~\cite{LIGOScientific:2016aoc}, the LIGO–Virgo–KAGRA Collaboration has observed more than two hundred events~\cite{LIGOScientific:2018mvr,LIGOScientific:2020ibl,LIGOScientific:2021usb,KAGRA:2021vkt,2025arXiv250818082T}. 
One of the most recent detections, GW250114, marks the onset of high signal-to-noise-ratio (SNR) BH spectroscopy~\cite{2025arXiv250908099T}. 
Such events allow for precise tests of general relativity (GR) in the strong-field regime by studying the ringdown phase of binary BH mergers~\cite{Berti:2025hly}. 
The latter describes the evolution of a perturbed remnant BH emitting characteristic GWs containing quasinormal modes (QNMs); see Refs.~\cite{Kokkotas:1999bd,Nollert:1999ji,Berti:2009kk,Konoplya:2011qq,Franchini2024,Berti:2025hly} for reviews on this topic.

Analyzing and interpreting such events requires the accurate computation of the QNM spectrum, which encodes rich information about BH properties and our understanding of GR, i.e., whether the final BH is described by the Kerr metric~\cite{Kerr:1963ud}. 
Current and next generation of GW detectors like the Einstein Telescope~\cite{ET:2025xjr} or LISA~\cite{LISA:2024hlh} will enable increasingly precise measurements of QNMs~\cite{Berti:2015itd,Barausse:2020rsu} and thus provide robust and stringent probes of gravity. 
However, to turn these measurements into insightful tests of GR and constrain modified theories of gravity, it is essential to obtain accurate BH QNM spectra beyond GR. 

Astrophysical binary BH mergers typically yield a spinning BH remnant, for which accurate QNM computations are significantly more challenging than for nonrotating BHs~\cite{Regge:1957td,Zerilli:1970se}. 
In GR, linear perturbations of the Kerr BH can be described by the Teukolsky equation~\cite{Teukolsky1972}, but similar calculations for spinning BHs in theories beyond GR have only been performed recently in a few cases.
Such studies rely on the slow-spin expansion, e.g., Refs.~\cite{Pierini:2021jxd,Wagle:2021tam,Srivastava:2021imr,Pierini:2022eim,Alapati:2025jmd}, or the small-coupling expansion, e.g., Refs.~\cite{Cano:2023tmv,Wagle:2023fwl,Cano:2024bhh,Cano:2024ezp}. 
Spectral methods that avoid the need for a simplified master equation are promising alternatives~\cite{Chung:2024vaf,Chung:2024ira,Blazquez-Salcedo:2024oek,Khoo:2024agm} and systematic extensions of the Teukolsky equation have been proposed in Refs.~\cite{Li:2022pcy,Hussain:2022ins,Cano:2023tmv,Cano:2024jkd}.

Starting from the decoupled master equation for linear perturbations, a variety of techniques have been developed to compute BH QNMs. 
They can be broadly classified into \emph{numerical and semi-analytical approaches}. 
Numerical methods, such as Leaver's continued fraction method, are generally highly accurate~\cite{Leaver1985}, while semi-analytical treatments provide physical insight into the underlying dynamics~\cite{Schutz:1985km}.
Among the numerical strategies, time-domain integration of the perturbation equation has a key role~\cite{Vishveshwara:1970zz}, as it allows to probe early and late time effects~\cite{Price:1971fb}, and can be generalized more easily to theories beyond GR~\cite{Doneva:2020nbb,Konoplya:2025afm}.
 
Among the semi-analytical methods, there are those that relate to key geometrical features of the spacetime, such as the photon sphere. 
Two such frameworks are the eikonal and post-Kerr~\cite{1972ApJ...172L..95G,Ferrari:1984zz,Glampedakis:2017dvb} methods. 
The eikonal one is particularly effective in the large-angular-momentum regime, although it can be extended beyond this limit. 
The post-Kerr one may be viewed as an application of the eikonal method to spacetimes that differ perturbatively from Kerr by a small deviation. 
This yields QNM frequencies expressed as power-series expansions in the small perturbation parameter.

In this paper, we benchmark the accuracy of different methods to compute QNMs of rotating BHs (time evolution, post-Kerr, and eikonal), and discuss their reliability for different ringdown SNRs. 
As a tractable problem and proof of principle, we study scalar field perturbations on the background of a modified Kerr BH~\cite{Konoplya:2018arm}, whose deformation affects the near-horizon structure, while preserving the asymptotic behavior. 
We extend a numerical code used in Refs.~\cite{Pedrotti:2023jhz,Pedrotti:2024znu} to compute 2+1-dimensional time-domain evolutions of scattering a Gaussian wave packet with the BH to excite QNMs for selected values of spin, beyond-Kerr deviation parameter, and multipoles.  
The simulations are analyzed via the \emph{Prony method}, a signal-processing technique that fits the ringdown waveform as a sum of damped sinusoids to directly infer the complex QNM frequencies~\cite{Berti2007}. 
We obtain accurate estimates of the fundamental mode $(n=0)$ for different multipoles $(\ell, m=\pm \ell)$, where the sign of $m$ distinguishes prograde and retrograde modes.

We then compare the numerical QNMs with predictions from the eikonal~\cite{Dolan:2010wr} and post-Kerr~\cite{Glampedakis:2019dqh} methods at different perturbative orders to assess their accuracy in capturing deviations from GR, i.e., the \emph{systematic errors}. 
Whether a given systematic error is acceptable for data-analysis applications depends on the expected \emph{statistical error}, which decreases with increasing ringdown SNR. 
The two types of uncertainties are then combined in the ``bias ratio'', i.e., the ratio between the systematic and statistical error. 
We use it as a criterion to assess the validity of the post-Kerr/eikonal approximation for a given QNM and estimate the maximum SNR until the QNM methods yield unbiased results. 

We find that the perturbation predominantly affects prograde modes, rather than retrograde ones. 
Moreover, the eikonal approximation accurately captures the nonlinear dependence of the fundamental modes on the perturbation parameter and, in several cases, is consistent with the Prony estimates within the method's uncertainties. 
By contrast, the post-Kerr expansion performs poorly at linear order and high spin, unless the deformation is very small. 
We also show that higher orders in the post-Kerr approximation can be rearranged via the Padé resummation~\cite{Basdevant1972,DelPiano2024, delpiano2025p}, which provides a remarkable improvement for large spin and perturbation parameter. 

The rest of this work is structured as follows. In Sec.~\ref{sec:geometry} we outline the geometrical setting, whereas in Sec.~\ref{sec:meth}, we describe the  methodology for QNM computations. 
Applications and results can be found in Sec.~\ref{sec:app_results}. Finally, we draw the conclusions in Sec.~\ref{sec:conc}. 
Throughout this work, we adopt units in which $G=c=1$.

\section{Geometrical setting}\label{sec:geometry}

In the following, we provide the mathematical details of the modified Kerr spacetime in Sec.~\ref{sec:modified-Kerr} and on the photon ring radius in Sec.~\ref{sec:photon-ring}. 
The modified scalar field perturbation equation is introduced in Sec.~\ref{sec:scalar-perturbation}.

\subsection{Modified Kerr black hole spacetime}\label{sec:modified-Kerr}

In the literature, there exist many proposals for theory-agnostic modifications of the Kerr metric developed with different motivations (see, e.g., Refs.~\cite{Johannsen:2013vgc,Johannsen:2011dh,Konoplya:2016jvv}, for some examples). 
In our work, we choose the modified Kerr BH originally introduced in Ref.~\cite{Caspar:2012ux} in the form adopted in Ref.~\cite{Schonenbach:2013nya,Konoplya:2016pmh} to study scalar QNMs, with a restricted choice allowing for separability~\cite{Konoplya:2018arm}. 
It describes a single-parameter deformation of the Kerr metric obtained via the substitution
\begin{equation}\label{deformation}
     M \to M+\frac{\epsilon}{2r^2}\,,
\end{equation}
where $\epsilon$ is the deformation parameter having dimensions of $M^3$, where, in the following, we identify $\epsilon$ with $\epsilon/M^3$. This parameter geometrically alters the relation between the BH mass $M$ and the position of the event horizon, while the large distance behavior is unaffected. 

This option is by no means unique, as there are countless ways of modifying the Kerr metric. However, our choice is primarily motivated by the following reasons: (1) symmetry arguments due to the separability of the perturbation equation~\cite{Konoplya:2018arm,pappas2018}; (2) possibility to give some insights on near horizon modifications, where possible new physics effects are expected to be relevant~\cite{Barman2021}; (3) interesting from the perspective of superradiance of scalar and electromagnetic perturbations~\cite{Franzin2021}.

In Boyer-Lindquist coordinates, the modified Kerr metric has the form of the Kerr spacetime, namely
\begin{align}\label{ModKerr_metric}
    \dd s^2&=-\left(\frac{\Delta-a^2\sin^2\theta}{\Sigma}\right)\dd t^2+\frac{\Sigma}{\Delta}\dd r^2-\frac{\sigma}{\Sigma}\sin^2\theta \,\dd\phi^2 \nonumber\\ 
    &+\Sigma\, \dd\theta^2-2a\sin^2\theta \left(\frac{r^2+a^2-\Delta}{\Sigma}\right)\dd t \dd\phi,
\end{align}
where $\Sigma=r^2+a^2\cos^2\theta$, $\Delta=r^2-2Mr+a^2-\epsilon/r$, and $\sigma=-(a^2+r^2)^2+a^2 \Delta\sin^2{\theta}$. The event horizons coincide with the Killing horizons and satisfy this equation in the equatorial plane placed at $\theta=\pi/2$
    \begin{equation}
        g^{rr}=\dfrac{\Delta}{\Sigma}=\frac{r^2-2Mr+a^2}{r^2}-\frac{\epsilon}{r^3}=0.
    \end{equation}
The solutions of this equation can be expressed as~\cite{Barman2021}
\begin{subequations}\label{horizons}
    \begin{align}
        r_H &= \frac{2M}{3}-\frac{\sqrt[3]{2}\,(3a^2-4M^2)}{3\sqrt[3]{\mathcal{A}}}+\frac{\sqrt[3]{\mathcal{A}}}{3\sqrt[3]{2}}\,,\\
        r_1 &= \frac{2M}{3}+\frac{(1+i\sqrt{3})(3a^2-4M^2)}{3   \sqrt[3]{4\,\mathcal{A}}}-\frac{(1-i\sqrt{3})\sqrt[3]{\mathcal{A}}}{6\sqrt[3]{2}}\,,\\
        r_2 &=\frac{2M}{3}+\frac{(1-i\sqrt{3})(3a^2-4M^2)}{3\sqrt[3]{4\,\mathcal{A}}}-\frac{(1+i\sqrt{3})\sqrt[3]{\mathcal{A}}}{6\sqrt[3]{2}}\,,
    \end{align}
\end{subequations}    
where
    \begin{align}
        \mathcal{A}&= 3\sqrt{3}\,(4a^6-4a^4M^2-36a^2M\epsilon+32 M^3\epsilon+27\epsilon^2)^{1/2} \nonumber\\
        &-18a^2 M + 16 M^3+27\epsilon\,.
    \end{align}
Depending on the values of $(\epsilon,a,M)$, $\mathcal{A}$ can take positive, negative, or complex values, thus Eqs. \eqref{horizons} can admit from no positive solutions (naked singularity) up to three positive solutions (where the event horizon is defined as the outermost one). 
Moreover, there is no upper bound $|a|\leq M$ on the spin parameter.

\subsection{Photon ring radius}\label{sec:photon-ring}

\begin{figure}
\includegraphics[width=1.0\linewidth]{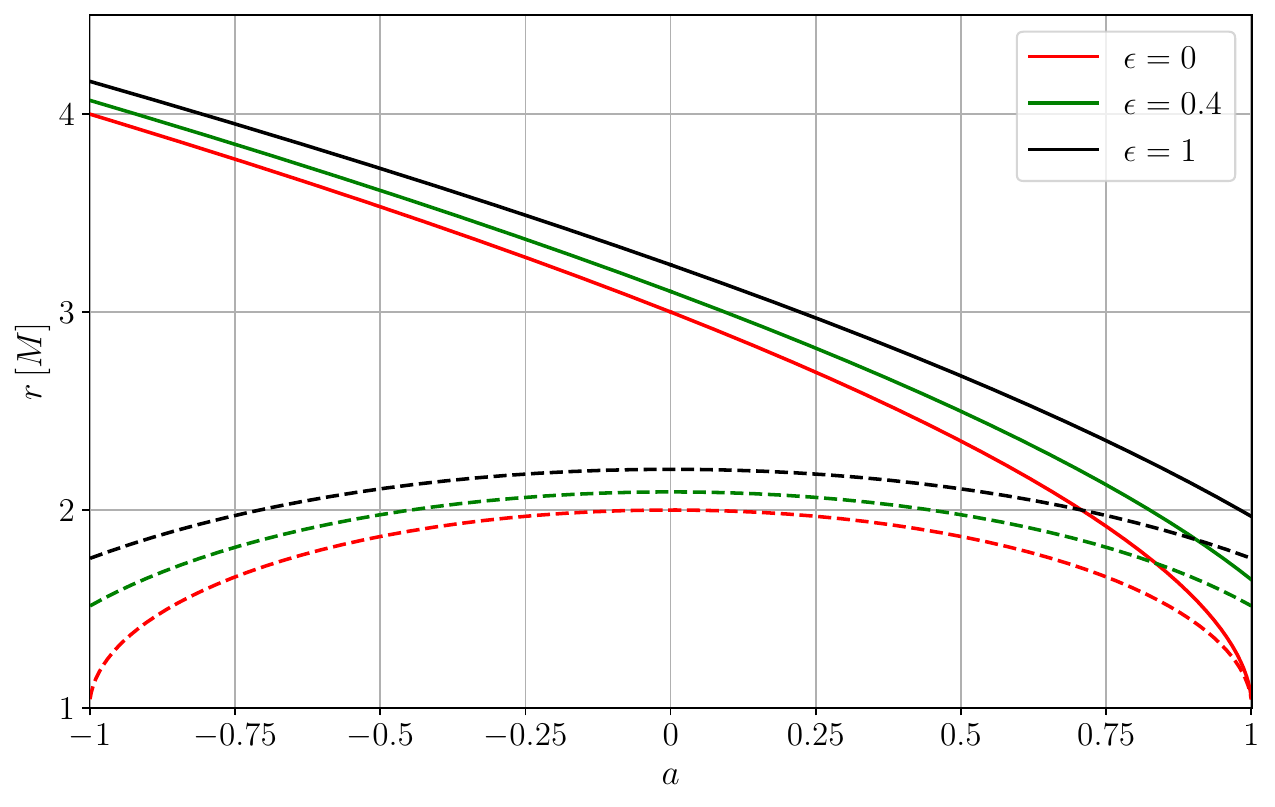}
\caption{The solid lines correspond to the prograde ($a>0$) and retrograde ($a<0$) equatorial photon orbits of the Kerr BH ($\epsilon=0$) and the modified Kerr BH ($\epsilon=0.4,1$). 
The dashed lines denote the position of the outer event horizons \eqref{horizons} for the same values of $\epsilon$.}
\label{Fig: photonspheres}
\end{figure}

Since the eikonal QNMs are associated with the properties of the photon sphere, it is useful to show the pertaining equation in the modified Kerr BH\footnote{Thanks to the separability argument~\cite{Konoplya:2018arm}, holding also in this metric, we are able to determine the null geodesic equation~\eqref{eq:null-geodesic}.}
    \begin{equation}\label{eq:null-geodesic}
        6Mr^4-2r^5+5r^2\epsilon\mp2ar^4\sqrt{4Mr+\epsilon/r}=0\,,
    \end{equation}
here and in what follows, the upper (lower) sign in $\mp$ will always correspond to prograde (retrograde) orbits, respectively. 
Analogously, the $\mp$ can also be absorbed in the sign of $a$, thus prograde (retrograde) quantities correspond to positive (negative) spin. The qualitative behavior of the equatorial photon orbits (and thus of the eikonal QNMs) can be understood by following this argument: prograde photon orbits lie closer to the event horizon than their retrograde counterparts at fixed spin. 
In addition, their radius decreases monotonically with increasing spin, whereas the retrograde radius exhibits the opposite trend, as illustrated in Fig.~\ref{Fig: photonspheres}. 
In the high-spin regime, prograde trajectories approach the horizon, thus probing an extremely near-horizon region, where the deformation is most pronounced. 
Consequently, the largest departures from the Kerr QNM spectrum are expected for high-spin prograde modes, while the smallest effects occur in the retrograde sector at large spin.

\subsection{Modified scalar perturbation equations}\label{sec:scalar-perturbation}

The modified Kerr metric belongs to the class of spacetimes that admit the separability of $\Phi$~\cite{Konoplya:2018arm} in the Klein-Gordon (KG) equation $\Box\,\Phi=0$ for a massless scalar field, also known as \emph{scalar Teukolsky equation}~\cite{Teukolsky1972}. Moreover, it admits separability of the Hamilton-Jacobi equation. This implies that there exists a Killing-Yano tensor and the associated Carter constant~\cite{Papadopoulos2018}. 

To obtain the KG equation in the time domain, the scalar field is decomposed as $\Phi(t,r,\theta,\varphi)=e^{im\varphi}\,\Psi(t,r,\theta)$, exploiting the axisymmetry of the background. As for the Kerr BH~\cite{Krivan1996}, the time evolution of the scalar field in Boyer-Lindquist coordinates can exhibit a pathological behavior near the horizon. For this reason, it is necessary to make this change of coordinates
    \begin{equation}
        \dd\tilde\phi=\dd\phi+\frac{a}{\Delta}\dd r\,, \;\;\;\;\;\;  \textrm{and} \;\;\;\;\;\; \frac{\dd r_*}{\dd r}=\frac{r^2+a^2}{\Delta}\,,
    \end{equation}
where $\tilde\phi$ is the equivalent of the Kerr azimuthal coordinate, while $r_*$ is the tortoise coordinate for the modified Kerr BH. In this new set of coordinates, the scalar field equation is given by
\begin{align}\label{eq_ModKerr}
    &\frac{\partial^2 \Psi}{\partial t^2}+\frac{(a^2+r^2)^2}{\sigma}\frac{\partial^2 \Psi}{\partial r_*^2}+\frac{\Delta}{\sigma}\frac{\partial^2 \Psi}{\partial \theta^2}-\frac{2iam(2Mr^2+\epsilon)}{r\sigma}\frac{\partial \Psi}{\partial t}\notag\\
    &+\frac{2iam(a^2+r^2)-2r\Delta}{\sigma} \frac{\partial\Psi}{\partial r_*}+\frac{\Delta}{\sigma}\cot{\theta}\frac{\partial \Psi}{\partial \theta}\notag\\
    &-\frac{m^2 \Delta}{\sigma\sin^2\theta}\Psi=0\,,
\end{align}
which is compatible with the Kerr case in the limit $\epsilon\to 0$~\cite{Krivan1996}. 
We focus on the perturbations of a massless scalar test field, since those can be studied self-consistently regardless of the underlying theory of gravity. 
A generalization to gravitational perturbation requires the knowledge of the field equations. In~\cite{Suvorov2020}, a theory admitting a modified Kerr BH as a vacuum solution within the framework of scalar-tensor theories has been constructed, and the corresponding gravitational QNMs of the non-rotating case were computed in Ref.~\cite{Suvorov:2021amy}. 
Moreover, Ref.~\cite{Franzin2021} investigated the type of matter content that would be required in GR to admit the modified Kerr BH as a solution. 
Both approaches can be used, in principle, as a starting point to investigate the gravitational perturbations of a modified Kerr BH.

Naturally, the modified Kerr BH spectrum does not depend only on mass and spin, but also on the additional parameter $\epsilon$, as the no-hair theorem does not hold~\cite{Cardoso:2016ryw}.

\section{Quasinormal mode methods}\label{sec:meth}

In the following, we outline methods for the computation of QNMs. 
In Sec.~\ref{sec:numerical-evolution}, we describe in detail how to numerically solve Eq.~\eqref{eq_ModKerr}. 
Then, Secs.~\ref{sec:eikonal} and~\ref{sec:post-Kerr} present the eikonal and post-Kerr approximations, respectively. 
Finally, in Sec.~\ref{sec:statistical_analysis} we outline the statistical analysis based on the bias ratio.

\subsection{Numerical evolution and QNM extraction}\label{sec:numerical-evolution}

\begin{figure}
\includegraphics[width=1.0\linewidth]{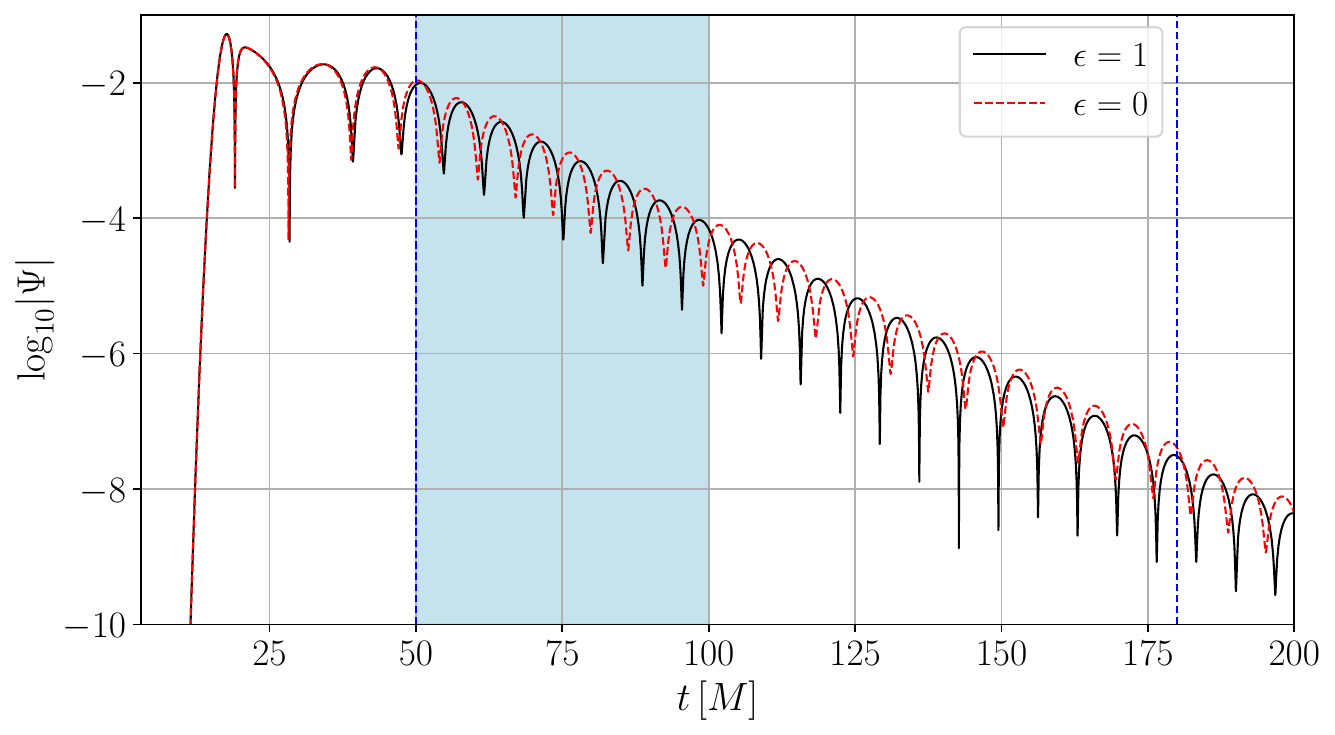}
\caption{Waveforms obtained setting $\ell=m=2$ and $a=0.3$ for Kerr ($\epsilon=0$) and modified Kerr BHs ($\epsilon=1$). The vertical blue dashed lines identify the starting $t_i=50\,M$ and ending $t_f=180\,M$ times. The shaded region denotes the range of starting times used in the Prony method.} 
\label{fig: waveform}
\end{figure}

Equation \eqref{eq_ModKerr} is solved numerically using the fourth-order Runge-Kutta method (see Appendix~\ref{sec:appendix} for details). 
Since we are interested in eikonal QNMs, we evolve the initial data in the range $\ell=m \in [2,10]$ up to times $t\gtrsim 200 M$. 
For each value of $\ell$, we consider four positive values for the perturbation parameter $\epsilon \in [0.1,0.2,0.4,1]$. 
We choose only positive values because the simulations exhibit an instability for negative $\epsilon$, which becomes more prominent for large spins. 
We also take two values for the spin: a small one $a=0.3$, and a high one $a=0.7$, being more interesting from an observational perspective~\cite{Fishbach2017}. 
Two waveforms for $\ell=m=2$ and $a=0.3$ are shown in Fig.~\ref{fig: waveform}, which appear to be dominated by the prograde mode of a given $(\ell,m)$ and its retrograde counterpart $(\ell,-m)$.

The QNM frequencies, as well as the amplitudes and phases of the modes, have been extracted using the Prony method~\cite{Berti2007}, which consists of fitting the waveform with a sum of damped sinusoids. 
To extract accurate QNM estimates, the Prony method must be applied only in the time interval $[t_i,t_f]$ where the signal is dominated by the QNM ringing, reducing the effect of the initial data at early times~\cite{Zhu:2024rej} and late-time tail~\cite{ DeAmicis:2024eoy}. 
This is shown in Fig.~\ref{fig: waveform}, where the vertical dashed blue lines correspond to the earliest possible starting time and the fixed ending time. 
The shaded region indicates the range of possible starting times $t_i$. 
Note that the waveforms for $\epsilon=0$ and $\epsilon=1$ are very similar at early times, with relevant differences only in the ringing phase. 
We find that the tail contribution becomes prominent only for $t > 200M$, and the simulations are dominated by the QNM ringing from $t \approx 50M$. 
For this reason, we set $t_f\approx180\,M$ around the end of the simulation time, and only vary the starting time $t_i$. 
This choice provides accurate estimates for the fundamental mode, since possible overtones have smaller damping times and only contribute to the early phase of the QNM ringing. 
The accuracy of this procedure is also discussed in Appendix~\ref{sec:appendix}.

\subsection{Eikonal approximation}\label{sec:eikonal}

In GR, the eikonal approximation~\cite{1972ApJ...172L..95G,Ferrari:1984zz,Dolan:2010wr,Yang:2012he} provides accurate estimates for QNMs with large angular momentum or multipole index $\ell \gg 1$. 
Under suitable assumptions~\cite{Cardoso:2008bp,Konoplya:2017wot}, eikonal QNMs can be directly related to two properties of the photon sphere around the BH~\cite{Pedrotti:2024znu}: \emph{orbital frequency} $\Omega$ (or angular velocity) and \emph{Lyapunov exponent} $\lambda$ (associated with the linear degree of instability). 

In our work, we are interested in the fundamental modes ($n=0$), satisfying also the condition $\ell=|m|$. 
The reason is twofold: (1) those are typically the most excited modes for gravitational perturbations; (2) they are associated with prograde and retrograde equatorial photon orbits~\cite{Glampedakis:2017dvb}. 
The fundamental QNM frequency $\omega$ can thus be expressed as
    \begin{equation}\label{eikonal_QNMs}
        \omega=\ell\,\Omega-\frac{i}{2}|\gamma|\,,
    \end{equation}
where
\begin{subequations}\label{eikonal_quantites}
    \begin{align}
        \Omega&=\frac{g'_{tt}}{-g'_{t\phi}\mp W^{1/2}}\,, \label{eq:eik_frequency}\\
        \gamma&=\Omega \,\sqrt{\frac{(g_{t\phi}^2-g_{tt}g_{\phi\phi})(g''_{tt} b^2+2g''_{t\phi} b+g''_{\phi\phi})}{2g_{rr}(g_{tt} b+g_{t\phi})^2}} \label{eq:eik_Lyapunov}\,.
    \end{align}
\end{subequations}    
Here $W=(g'_{t\phi})^2-g'_{tt}g'_{\phi\phi}$ and $b=\Omega^{-1}$ is the impact parameter of the photon sphere. 
The primes denote the derivatives with respect to the radial coordinate, and all quantities are evaluated at the photon sphere radius. 
At this point, it is worth noticing that the identification with QNMs and geodesics is an assumption that may not hold in general theories beyond GR or alternative to it, as metric-affine theories of gravity~\cite{Faraoni:2010pgm, Capozziello:2022zzh}, so this is another layer of "approximation". 
See also Refs.~\cite{Khanna:2016yow,Konoplya:2017wot,Konoplya:2022gjp} for specific studies on this topic.

\subsection{Post-Kerr approximation}\label{sec:post-Kerr}

An alternative method to compute the eikonal QNMs of an axisymmetric BH spacetime is the so-called post-Kerr approximation~\cite{Glampedakis:2017dvb}. 
This approach is valid for spacetimes that deviate from Kerr only by a small amount, namely, the metric tensor can be written as
    \begin{equation}\label{PKmetric}
        g_{\mu\nu}=g^{\rm Kerr}_{\mu\nu}+\epsilon \,h_{\mu\nu}+\mathcal{O}(\epsilon^2),
    \end{equation}
where $|\epsilon\,h_{\mu\nu}| \ll1$ in such a way that the deformation on the Kerr observable can be considered small. The post-Kerr estimates for the photon sphere position $r_{\rm ph}$, the impact parameter $b$, the orbital frequency $\Omega$, and the Lyapunov exponent $\lambda$ are then obtained as a power series in $\epsilon$ 
\begin{subequations}\label{postKerr_quantities}
\begin{align}
    r_{\rm ph}&=r^{\textrm{Kerr}}_{\rm ph}+\delta r_1 \,\epsilon + \delta r_2 \,\epsilon^2+\mathcal{O}(\epsilon^3)\,,\\
    b&=b^{\textrm{Kerr}}+\delta b_1 \,\epsilon + \delta b_2 \,\epsilon^2+\mathcal{O}(\epsilon^3)\,,\\
    \Omega&=\Omega^{\textrm{Kerr}}+\delta \Omega_1 \,\epsilon + \delta \Omega_2 \,\epsilon^2+\mathcal{O}(\epsilon^3)\,,\\
    \gamma&=\gamma^{\textrm{Kerr}}+\delta \gamma_1 \,\epsilon + \delta \gamma_2 \,\epsilon^2+\mathcal{O}(\epsilon^3)\,,
\end{align}
\end{subequations}
through a suitable set of coefficients $\{\delta r_i, \,\delta b_i,\, \delta \Omega_i,\, \delta \gamma_i\}$ due to the Taylor expansion.

This ansatz can be used to solve Eqs.~\eqref{eq:eik_frequency} and~\eqref{eq:eik_Lyapunov} order by order in $\epsilon$. 
For consistently computing the higher orders in the post-Kerr approximation, it is necessary to keep higher orders in $\epsilon$ also in the metric expansion Eq.~\eqref{PKmetric}. 
The explicit expressions for $\delta\Omega_1$ and $\delta\gamma_1$ can be found in~\cite{Glampedakis:2017dvb}. 
They are quite involved, but depend only on the derivatives of $h_{\mu\nu}$ up to the second order.

Next, we apply the formalism to the modified Kerr spacetime. 
From Eq.~\eqref{ModKerr_metric} we immediately see that the metric does not have the form Eq.~\eqref{PKmetric}. 
However, the metric coefficients in Eq.~\eqref{ModKerr_metric} can be Taylor-expanded at first order in $\epsilon$, and give specific expressions for the metric deformation $h_{\mu\nu}$. 
In addition to that, the deformation Eq.~\eqref{deformation} affects most of the metric coefficients linearly, except for $g_{rr}$, which depends in a non-linear way on $\epsilon$. 
This is relevant for the eikonal calculations, because $g_{rr}$ appears only in the definition of the Lyapunov exponent, not in the orbital frequency. 
This implies that the modified Kerr BH is basically a linearized metric for the real part of the QNM frequency and a non-linear metric for the imaginary part. 

Let us also highlight the main differences between the eikonal and post-Kerr methods. 
First, the eikonal approximation requires the observables $\{r_{\rm ph},b,\Omega,\lambda\}$ related to the generic axisymmetric spacetime under consideration, which, in most cases, can only be determined numerically. 
On the other hand, the post-Kerr approximation only needs the Kerr observables and the deformation $h_{\mu\nu}$, which are known analytically. 

Besides the post-Kerr predictions, we also compute the Padé approximants~\cite{Basdevant1972} of the third-order post-Kerr series. 
This consists of resumming the series as a rational function and typically provides a better and faster convergence than the Taylor series.

Moreover, in order to compare the eikonal (E) and post-Kerr (pK) estimates with the Prony ones, the former have to be first calibrated to the Kerr QNMs. 
This is achieved by defining the fundamental QNMs of the modified Kerr BH via the following formula
    \begin{equation}\label{calibration}
        \omega^{\textrm{E/pK}}(M,a,\epsilon) = \omega^\text{Kerr}(M,a) + \delta \omega^\text{E/pK} (M, a, \epsilon)\,,
    \end{equation}
where
    \begin{equation}
        \delta \omega^\text{E/pK} (M, a, \epsilon)
= \bar\omega^\text{E/pK}(M, a, \epsilon)- \bar\omega^\text{E/pK} (M, a,0)\,,
    \end{equation}
and the bar denotes the QNM estimates directly obtained from Eq.~\eqref{eikonal_QNMs} with the Eikonal Eq.~\eqref{eikonal_quantites} or post-Kerr methods Eq.~\eqref{postKerr_quantities}. 
This provides unbiased estimates in the GR limit, since the scalar Kerr QNMs can be computed with pristine accuracy and the errors from the eikonal/post-Kerr approximation only affect QNM shifts $\delta \omega^\text{E/pK}$. 
In order to obtain accurate estimates of the Kerr QNMs, we use the \texttt{Python} package \texttt{qnm}~\cite{Stein:2019mop} based on Leaver's method~\cite{Leaver1985}.

\subsection{Statistical analysis}\label{sec:statistical_analysis}

To determine the practical relevance of the theoretical errors of the post-Kerr and eikonal methods, we estimate the statistical errors on the fundamental QNM at different SNRs. 
This assessment is crucial for determining whether systematic (or theoretical) errors remain within tolerable limits for a given measurement, and thus for establishing the reliability of each approach. 
When systematic errors exceed statistical ones, parameter-estimation applications may yield unreliable results. 
For instance, the true QNM may fall outside the expected range of a posterior distribution, and one would conclude false bounds on deviations from GR. 

\begin{figure*}[t]
    \centering
    \includegraphics[width=\linewidth]{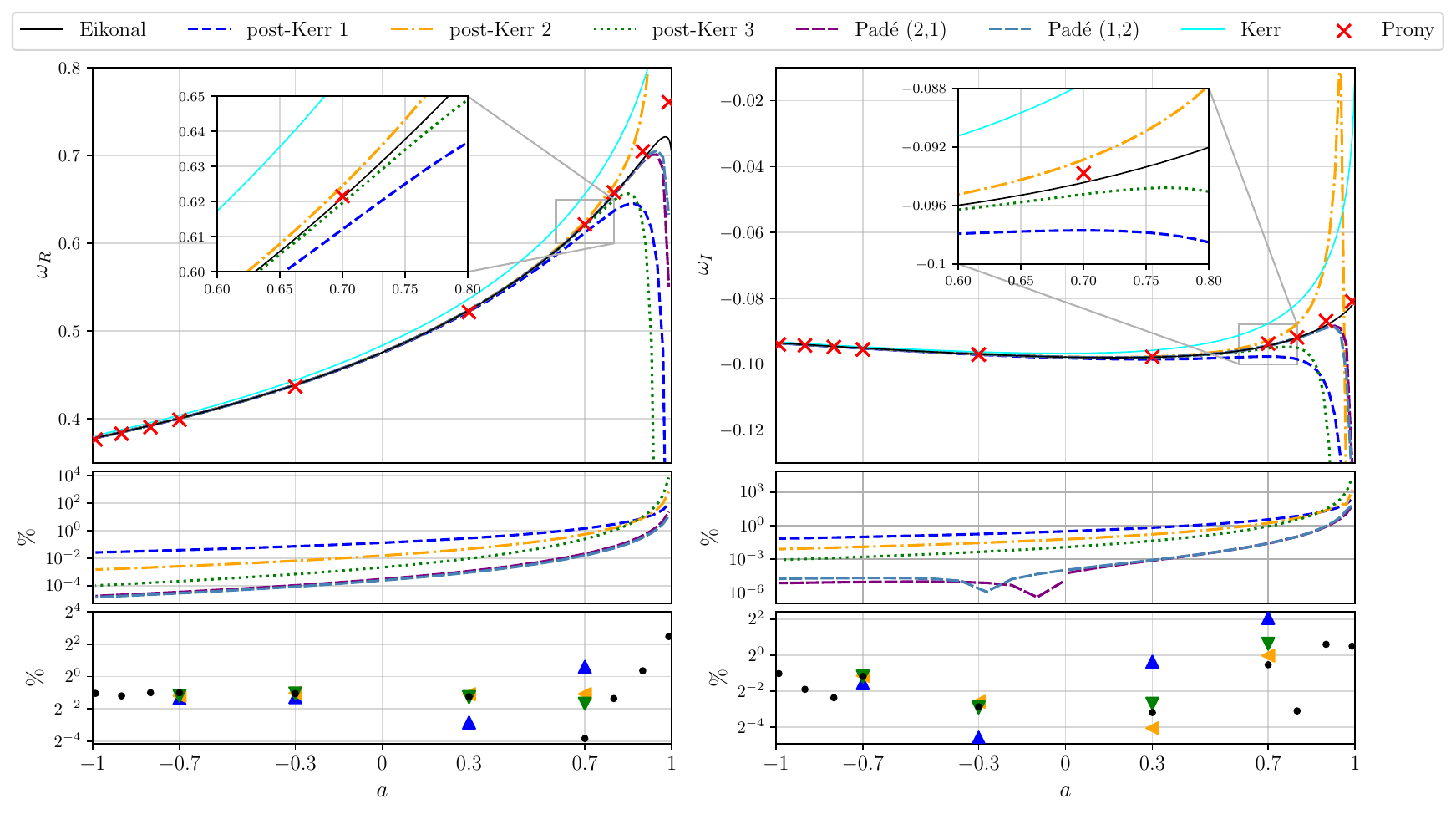}
    \caption{First row: estimates of the real and imaginary parts of prograde and retrograde modes as a function of the BH spin for $\ell=2$ and $\epsilon=0.4$ from the eikonal method, Prony extraction and different orders in the post-Kerr and Padé approximation. The Kerr values ($\epsilon=0$) are reported as a reference of the unperturbed case. Second row: relative error in $\log_{10}$ scale of post-Kerr and Padé approximations with respect to the eikonal predictions. Third row: relative error in $\log_2$ scale of the eikonal (black dots) and post-Kerr order 1 (blue triangles), post-Kerr order 2 (orange triangles), post-Kerr order 3 (green triangles) estimates with respect to the Prony ones. 
    The Padé estimates are not reported as they overlap with the eikonal ones. 
    }
    \label{fig:QNMs}
\end{figure*}

We estimate the statistical errors in a simple setup, because we are only interested in providing a rough estimate. 
We assume that a given ringdown signal is analyzed such that it can be well approximated by a single damped sinusoid. 
Furthermore, since we only consider a single mode, we also adopt the assumption of flat noise (see Appendix~\ref{sec:dataanalysis} and Ref.~\cite{Volkel:2025jdx} for a similar analysis and more details). 
A more realistic estimate would require a detector-specific power-spectral density, selecting the mass and spin of the final BH, and changing the number of modes in a ringdown signal or even providing a full time evolution of physical initial data. 
Addressing the previous points is clearly beyond our scope, as we are only interested in a quick and rough estimate of how the statistical errors of the fundamental mode parameters depend on the SNR. 

We introduce the so-called bias ratio $\mathcal{R}$ to express the importance of systematic and statistical errors. 
It is defined as the systematic error $\delta \omega$ (related to the selected approximation with respect to the Prony method) over the statistical error $\Delta \omega$
\begin{align}\label{eq:bias-ratio}
\mathcal{R} = \frac{\delta \omega}{\Delta \omega}\,.
\end{align}

\section{Applications and results}\label{sec:app_results}

In the following, we calculate the scalar field QNM frequencies in the deformed Kerr metric. 
First, we report the accuracy of the eikonal and post-Kerr approximations in Sec.~\ref{sec:prediction-power}. 
We then discuss the suitability of both methods to predict deviations from Kerr QNMs in Sec.~\ref{sec:eikonal}, and finally relate the theoretical errors to statistical errors at different SNRs in Sec.~\ref{sec:relation-SNR}.

\subsection{Accuracy of QNM methods}\label{sec:prediction-power}

\begin{figure*}
\includegraphics[width=1.0\linewidth]{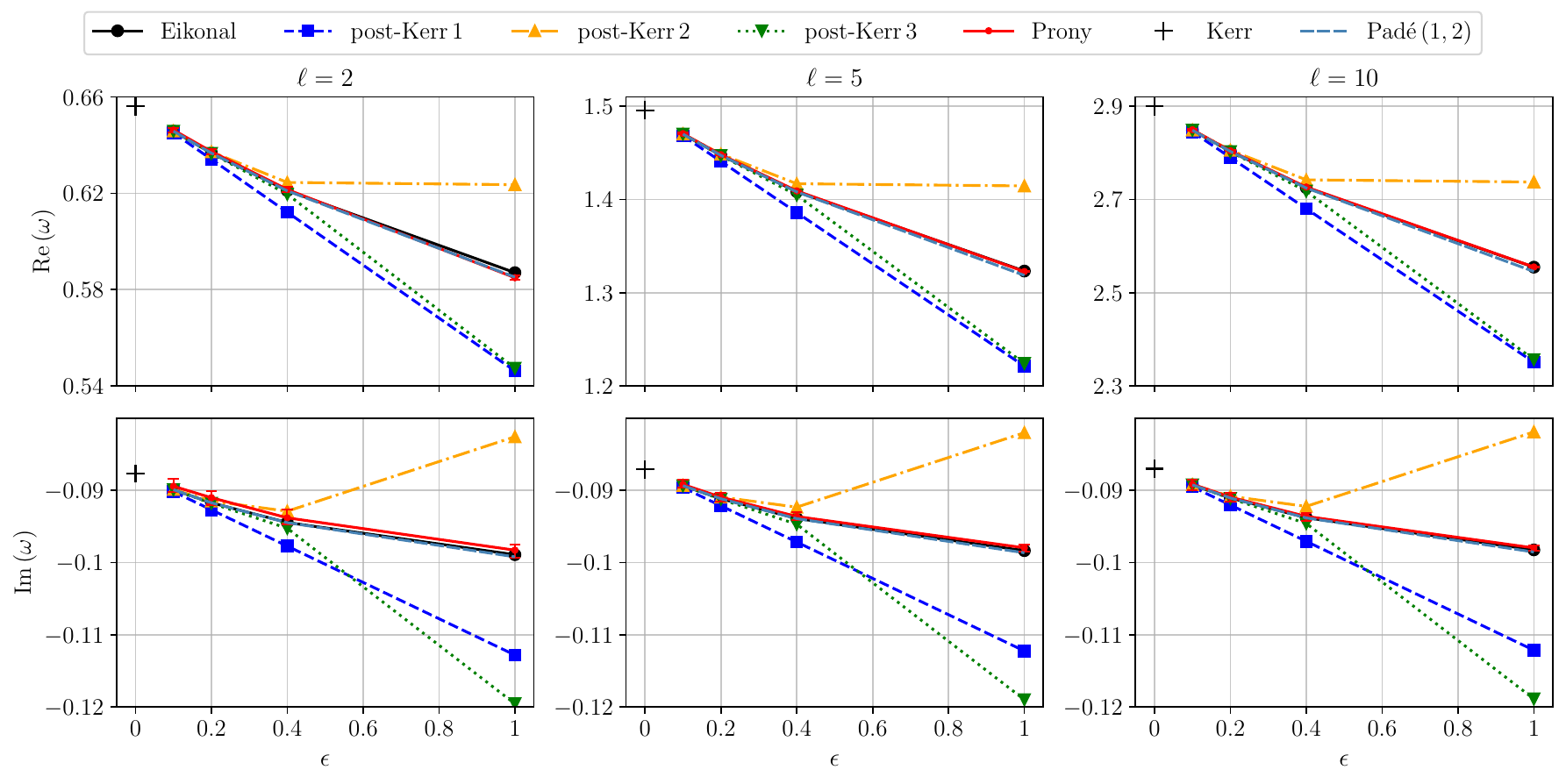}
\caption{Fundamental QNMs for different deviation parameters $\epsilon$ for selected values of $\ell$ with $m=\ell$ computed with eikonal (black dots), post-Kerr (blue squares), post-Kerr order 2 (orange triangles), post-Kerr order 3 (green triangles), Padé order $(1,2)$ (light blue) and Prony method (red dots) for spin $a=0.7$ for real (top panels) and imaginary (bottom panels) parts. 
The Kerr values are shown for comparison (black +). 
The estimation for Prony method errors is indicated by error bars (very small). 
The Padè estimates at order $(2,1)$ are not reported as they overlap with order $(1,2)$. 
} 
\label{fig:pK_orders_07}
\end{figure*}

Let us first discuss the results of the eikonal and post-Kerr approximations for the modified Kerr BH. 
In the top row of Fig.~\ref{fig:QNMs}, we report the prograde and retrograde fundamental modes for $\ell=2,\epsilon=0.4$ as a function of $a$ (see Fig.~\ref{fig:QNMs_l10} for the case $\ell=10,\epsilon=0.4$). 
Those plots show a qualitative difference from the Kerr case in both the real and imaginary parts ($\omega_R$ and $\omega_I$, respectively). 
In fact, $\omega_R$ grows with $a$ and has a maximum around $a\approx0.978$, while for the Kerr BH, $\omega_R$ grows monotonically with $a$. 
This is mainly due to the calibration procedure (see Eq.~\eqref{calibration}) which has a stronger effect on the low multipoles like $\ell=2$ where the eikonal approximation is less accurate. 
In fact, for $\ell=10$ (Fig.~\ref{fig:QNMs_l10}) this effect is less prominent since the eikonal approximation improves. 
On the other hand, Fig.~\ref{fig:QNMs} shows that $\omega_I$ does not change significantly over the entire range of spin. 
This is again quite different from the eikonal Kerr QNMs, where the imaginary part tends to approach zero for very large spin.

The red crosses in Fig.~\ref{fig:QNMs} correspond to the QNMs extracted using the Prony method. 
In addition to the ones at $a=0.3,0.7$, we here also provide results at $a=(0.8,0.9,0.99)$ to probe the large spin behavior. 
One can notice that for $a \gtrsim 0.9$ the eikonal approximation becomes significantly less accurate. Two effects may give rise to this result. 
On the one hand, the eikonal approximation is less accurate at large spin, especially at low values of $\ell$. 
On the other hand, the accuracy of the numerical simulations and Prony extraction may also degrade in this limit. 
Nevertheless, we have checked that increasing the radial and angular resolution up to $20\%$ of the original values, as well as changing the shape of the initial Gaussian profile, do not affect the convergence of the Prony method.

\begin{figure*}[t]
\includegraphics[width=0.95\linewidth]{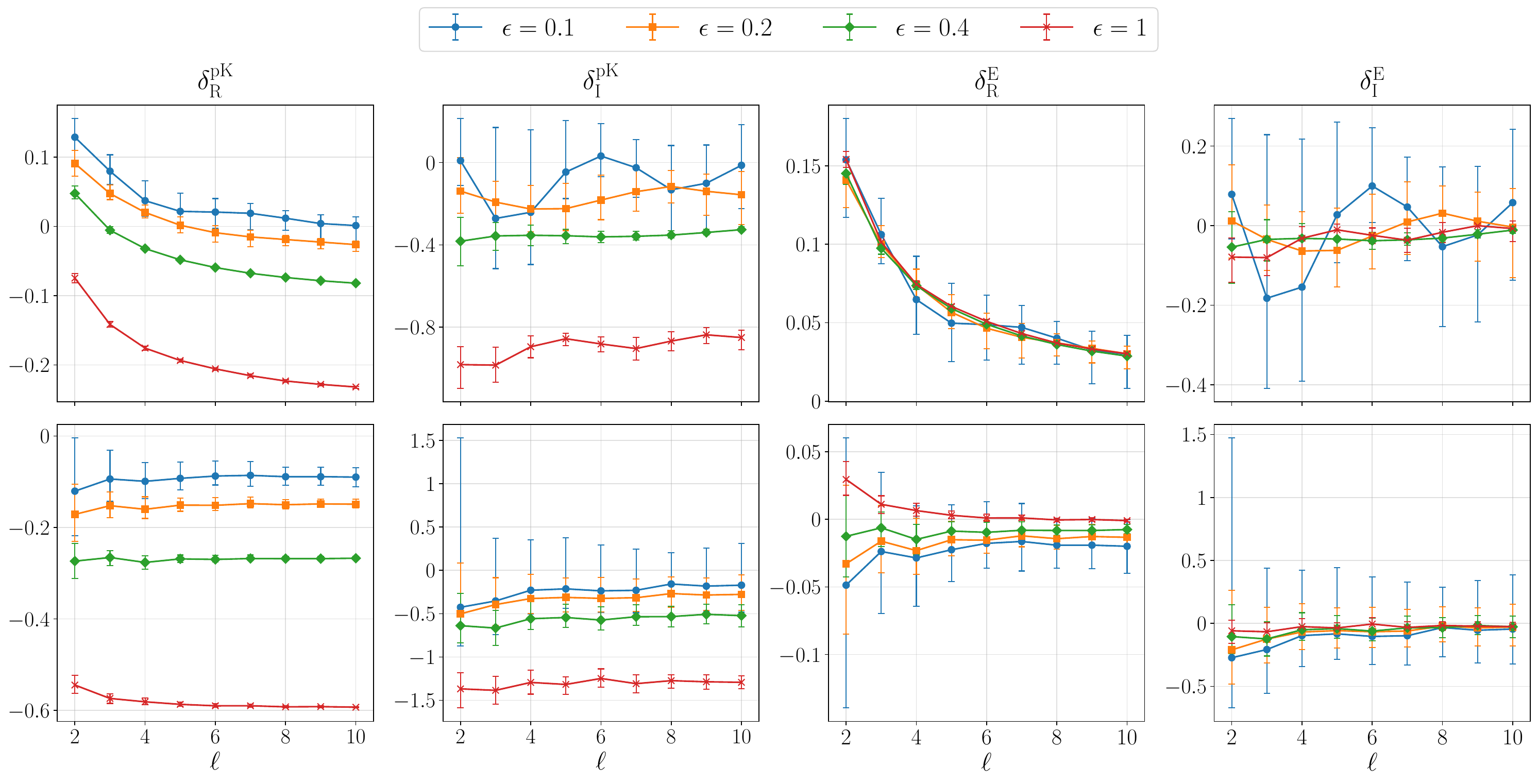}
\caption{Plots of $\delta^{\textrm{E/pK}}_R$ and $\delta^{\textrm{E/pK}}_I$ for a given $\epsilon$ as a function of $\ell$ for spins $a=0.3$ (top panels) and $a=0.7$ (bottom panels).
}
\label{Fig:comparison_combined}
\end{figure*}

Regarding the post-Kerr approximation, we note that the second order shows a qualitatively different behavior at large spin, overestimating the eikonal results in the real part, while peaking at values close to zero for the imaginary part. 
By computing even higher orders, we found that the fourth order exhibits a similar behavior, pointing toward a basic difference between even and odd orders of the post-Kerr approximation.

Figure~\ref{fig:QNMs} also shows that the convergence of the post-Kerr series at high spin is rather slow. For this reason, we report the results for Padé resummations of the third order post-Kerr series at order $(2,1)$ and $(1,2)$. 
They appear very similar over the entire range of spin, especially in the real part, and at both orders give much better results than the post-Kerr approximation. 
This improvement can be seen in the middle panels of Fig.~\ref{fig:QNMs}, where there are the relative errors between the post-Kerr/Padé approximations and the eikonal predictions. 
One can clearly see that higher orders $\epsilon$ in the post-Kerr provide remarkable improvement for both real and imaginary parts, especially for negative and low spins~\cite{Glampedakis:2017dvb}. 

Finally, we underline the relation between prograde and retrograde photon orbits and the corresponding QNMs. 
Bearing in mind Fig.~\ref{Fig: photonspheres}, the relative deviation with respect to the Kerr case can reach up to $12\%$ for the prograde modes and $3\%$ for the retrograde ones over the range of values of $\epsilon$ considered in this paper.

Figure~\ref{fig:QNMs} also offers a direct comparison between different post-Kerr orders. 
In particular, the bottom panels show that the three post-Kerr profiles have the same trend up to $a\sim0.8$. 
Below this threshold, the post-Kerr convergence is uniform, as any additional order improves the accuracy of the approximation. 
However, for spin larger than the threshold the situation is reversed, and higher orders do not lead to any improvement. 
In fact, in those regimes the relative errors become very large. 

Moreover, Fig.~\ref{fig:QNMs} is obtained for $\epsilon=0.4$, therefore it is important to study the behavior for different values of the perturbative parameter. 
It turns out that, as we increase $\epsilon$, the spin threshold decreases and the region of uniform convergence will shrink accordingly. 
This is expected as the post-Kerr approximation works better for smaller values of $\epsilon$. 
Moreover, these considerations apply mainly to prograde modes, as the retrograde ones are accurately estimated by higher orders even for $a\approx1$. 
This is expected from our discussion in Sec.~\ref{sec:photon-ring} on the photon rings, because the location of the retrograde one is less impacted by the near-horizon modification. 

The effect of the higher orders in the post-Kerr approximation is also reported in Fig.~\ref{fig:pK_orders_07} for the case $a=0.7$ and selected values of $\ell$ (the case $a=0.3$ is reported in the Appendix~\ref{sec:complementary results}). 
The figure indicates that, for small values of $\epsilon$, there is a remarkable improvement across all values of $\ell$, as the estimates converge to the eikonal and Prony results. 
For $\epsilon=1$, however, the picture is significantly different, and it is important to distinguish between spin $a=0.3$ and $a=0.7$. 
In fact, $a=0.7$ turns out to be close to the threshold for $\epsilon=1$, so there is no advantage in the higher post-Kerr orders. 
Figure~\ref{fig:pK_orders_07} shows that in most cases it is actually worse to rely on higher orders. 
Conversely, $a=0.3$ lies still well below the threshold even for $\epsilon=1$, and there is a clear improvement in the higher post-Kerr orders for all values of $\epsilon$.

\subsection{Suitability of eikonal and post-Kerr methods}\label{sec:suitability-eikonal}

The suitability of the eikonal or post-Kerr approximation can be measured by introducing the functions
\begin{subequations}\label{eq:comparison-function}
\begin{align} 
\delta^{\textrm{E/pK}}_R&=\frac{\omega^{\textrm{\textrm{Pr}}}_R-\omega^{\textrm{E/pK}}_R}{\omega^{\textrm{Pr}}_R-\omega^{\textrm{Kerr}}_R}\,,\label{delta_R}\\
\delta^{\textrm{E/pK}}_I&=\frac{\omega^{\textrm{\textrm{Pr}}}_I-\omega^{\textrm{E/pK}}_I}{\omega^{\textrm{Pr}}_I-\omega^{\textrm{Kerr}}_I}\,,\label{delta_I}
\end{align}
\end{subequations}
for real and imaginary parts of the QNMs obtained from the eikonal (E), post-Kerr (pK), and Prony (Pr) methods. 
The numerator contains the difference between the Prony and eikonal/post-Kerr estimates, i.e., their systematic error. 
Instead, the denominator compares the Prony estimates with the Kerr QNMs, thus quantifying the effect of the BH deformation on QNMs (hereafter denoted as "GR shift").

It is important to note that the Prony method is thus valid only if $\delta^{\textrm{E/pK}}_{R/I}$ is less than unity, i.e., the effect of the modified spacetime dominates over the accuracy of the Prony method. 
Notice also that, due to the calibration of the QNM estimates in Eq.~\eqref{calibration}, $\delta^{\textrm{E/pK}}_{R/I}$ is ill-defined for $\epsilon=0$ as the GR shift would be equal to zero.

In Fig.~\ref{Fig:comparison_combined}, we report $\delta^{\textrm{E/pK}}_R$ and $\delta^{\textrm{E/pK}}_I$ as a function of $\ell$ for all values of $\epsilon$ and spin considered in this paper. 
The error bars are obtained by propagating the errors on the Prony estimates. 
It is interesting to notice that in the eikonal case, the comparison function Eq.~\eqref{eq:comparison-function} is always smaller than one. 
Moreover, $\delta^{\textrm{E}}_R$ shows a decreasing behavior as a function of $\ell$, especially for $a=0.3$. 
This trend is expected since the real part of the eikonal QNMs scales with $\ell$ as in Eq. \eqref{eikonal_QNMs}. 
Therefore, in Eq. \eqref{delta_R} the GR shift is proportional to $\ell$. 
The systematic error, on the other hand, is also expected to decrease as a function of $\ell$, as the eikonal approximation becomes more accurate, but this effect is subdominant compared to the GR shift. 
The same argument can also be applied to $\delta^{\textrm{E}}_I$, since the imaginary part of the eikonal estimate is independent from $\ell$. 
This implies that the GR shift remains almost constant over $\ell$. Moreover, it will be quite small, especially for $\epsilon=0.1$, where the deviations from GR are minimal. 
This also explains the large error bars in $\delta^{\textrm{E}}_I$ for $\epsilon=0.1$ compared to the other cases. 
For the systematic error, it is still expected to decrease for larger $\ell$, but as in the previous case, this effect is not very strong over the values of $\ell$ considered in this paper.

Let us now turn to the first-order post-Kerr results. 
One can immediately notice that in the case $\epsilon=1$ and $a=0.7$, $\delta^{\textrm{pK}}_I$ is beyond unity. 
Also, for $a=0.3$ it is very close to the threshold, suggesting that for $\epsilon \approx1$ the post-Kerr approximation is not very effective. 
Moreover, the lines $\epsilon=1$ are well separated from the others. 
The behavior of the real part $\delta^{\textrm{pK}}_R$ is even more remarkable, as it appears to increase in absolute value as a function of $\epsilon$, except for $\epsilon=0.1$. 
This implies that, for $\epsilon > 0.1$, the systematic error dominates over the GR shift even for large $\ell$. 
The upshot of these considerations is that the first-order post-Kerr approximation does not behave as expected by the eikonal method and is limited to small deviations from Kerr. 
Similar plots can also be realized for the higher orders in the post-Kerr approximation, but they reveal to be not particularly informative.

\subsection{Relation to ringdown SNRs}\label{sec:relation-SNR}

The statistical error can be quantified by applying Bayesian statistics. As anticipated in Sec.~\ref{sec:statistical_analysis}, we inject the Prony estimates of each mode into a damped sinusoid, and then extract the parameters and their statistical uncertainty using Markov-chain Monte Carlo (MCMC) sampling. 
This procedure has been repeated for many values of the SNR $\rho\in[20,500]$. 

\begin{figure}[t]
\includegraphics[width=1.0\linewidth]{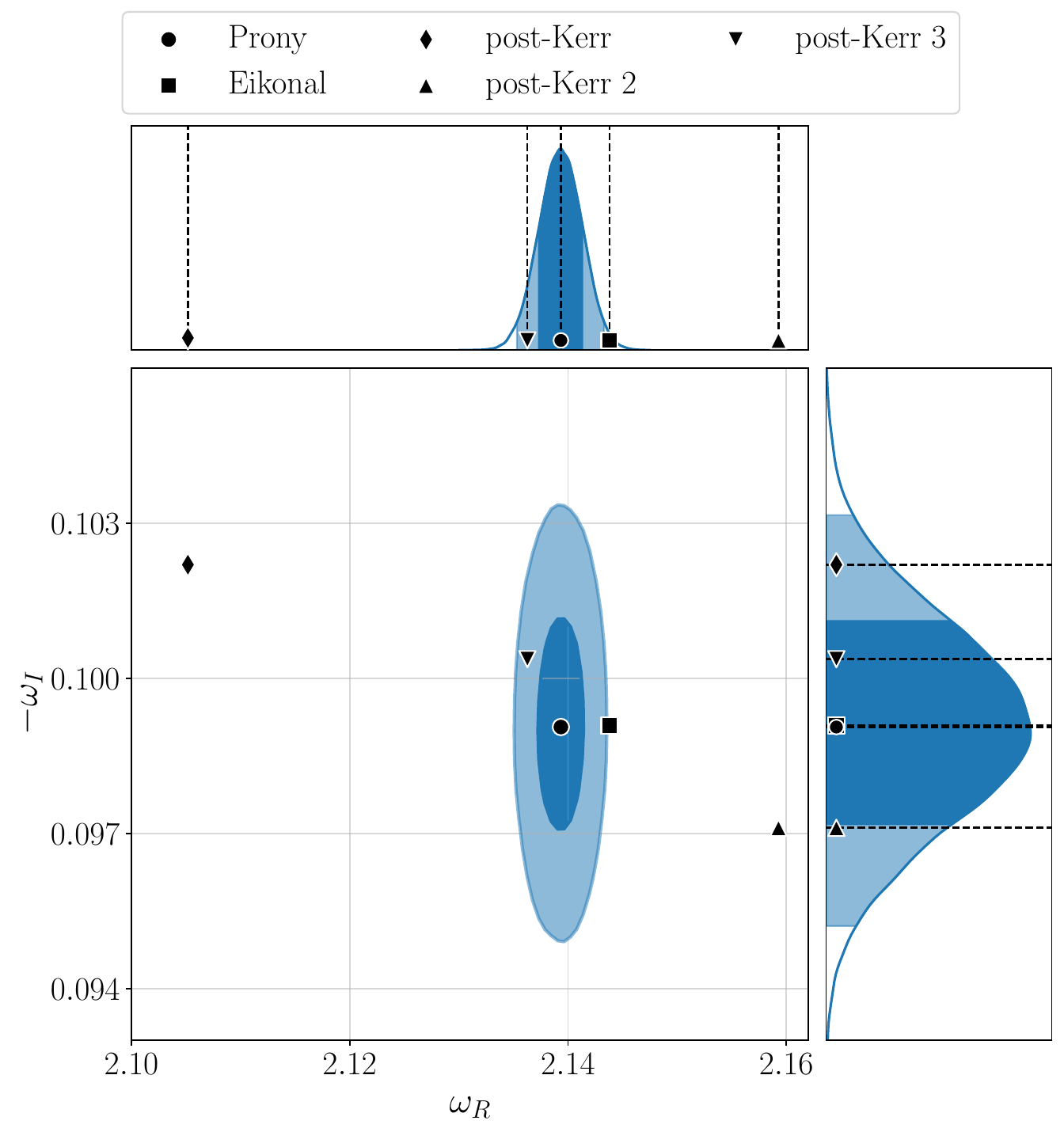}
\caption{Corner plot for $\omega_I$ vs $\omega_R$, obtained for the mode $(\ell=10, \,a=0.3,\,\epsilon=1)$ at SNR $\rho=100$, with the $1\sigma$ and $2\sigma$ confidence intervals. 
The  Kerr $(\epsilon=0)$ QNM $\omega^\textrm{Kerr}=2.28-0.095\,i$ is well outside the confidence interval.
}
\label{Fig: corner}
\end{figure}

Figure~\ref{Fig: corner} shows the corner plot obtained at $\rho=100$ for the mode $(\ell=10, a=0.3, \epsilon=1)$, where the Prony estimate, as well as the eikonal and post-Kerr predictions at different orders, are shown with the $1\sigma$ and $2\sigma$ confidence intervals. 
In this case, the eikonal prediction offers an accurate estimate of the Prony injection, especially for the imaginary part. 
While the first-order post-Kerr approximation is not very accurate, there is a significant improvement for higher orders. 
Furthermore, since larger SNR yields smaller statistical error, there exists a sufficiently high SNR such that even the eikonal and the third order post-Kerr approximation lie outside the $2\sigma$ confidence interval.

It turns out that, for all values of $\rho$ considered in this paper, the statistical error can be well approximated by its large SNR limit 
\begin{equation}\label{eq_stat_err_const}
    \Delta\omega = \frac{\bar{c}}{\rho}\,,
\end{equation}
where $\bar{c}\approx0.6$ is a constant (see Appendix~\ref{sec:dataanalysis}, for details). 
This implies that the statistical error is mainly fixed by the SNR $\rho$ and depends weakly on the other parameters.

Using Eq.~\eqref{eq_stat_err_const} and setting the bias ratio $\mathcal{R}=1$ then allows one to approximate the SNR $\rho$ at which the systematic error becomes dominant via
\begin{equation}
    \mathcal{R}\approx\frac{\rho\,\delta\omega}{\bar{c}}=1 \implies \rho=\frac{\bar{c}}{\delta\omega}\,,
\end{equation}
which is a simple formula for the SNR as a function of the systematic error. 
For each value of $(a,\ell,\epsilon)$, we apply this relation to compute the SNR at which the bias ratio threshold is reached. For any given spin and multipole, we find that the SNR threshold decreases with $\epsilon$, as larger $\epsilon$ correspond to higher deviations from GR. Moreover, we find that, considering both the real and imaginary part of the QNMs for $a=0.7, \ell=2, \epsilon=0.2$, the post-Kerr predictions up to third order become unreliable for $\rho \sim 200$, while for the eikonal prediction we find $\rho\sim1000$. On the other hand, for $a=0.3$ the post-Kerr predictions are very close to the eikonal ones and the bias threshold is reached at $\rho\sim600$. Combining the results for prograde and retrograde modes, we obtain more than one hundred SNR values for each method (eikonal, post-Kerr, Padé) and for the real and imaginary parts of the QNMs. 

\begin{figure*}[t]
\includegraphics[width=1.0\linewidth]{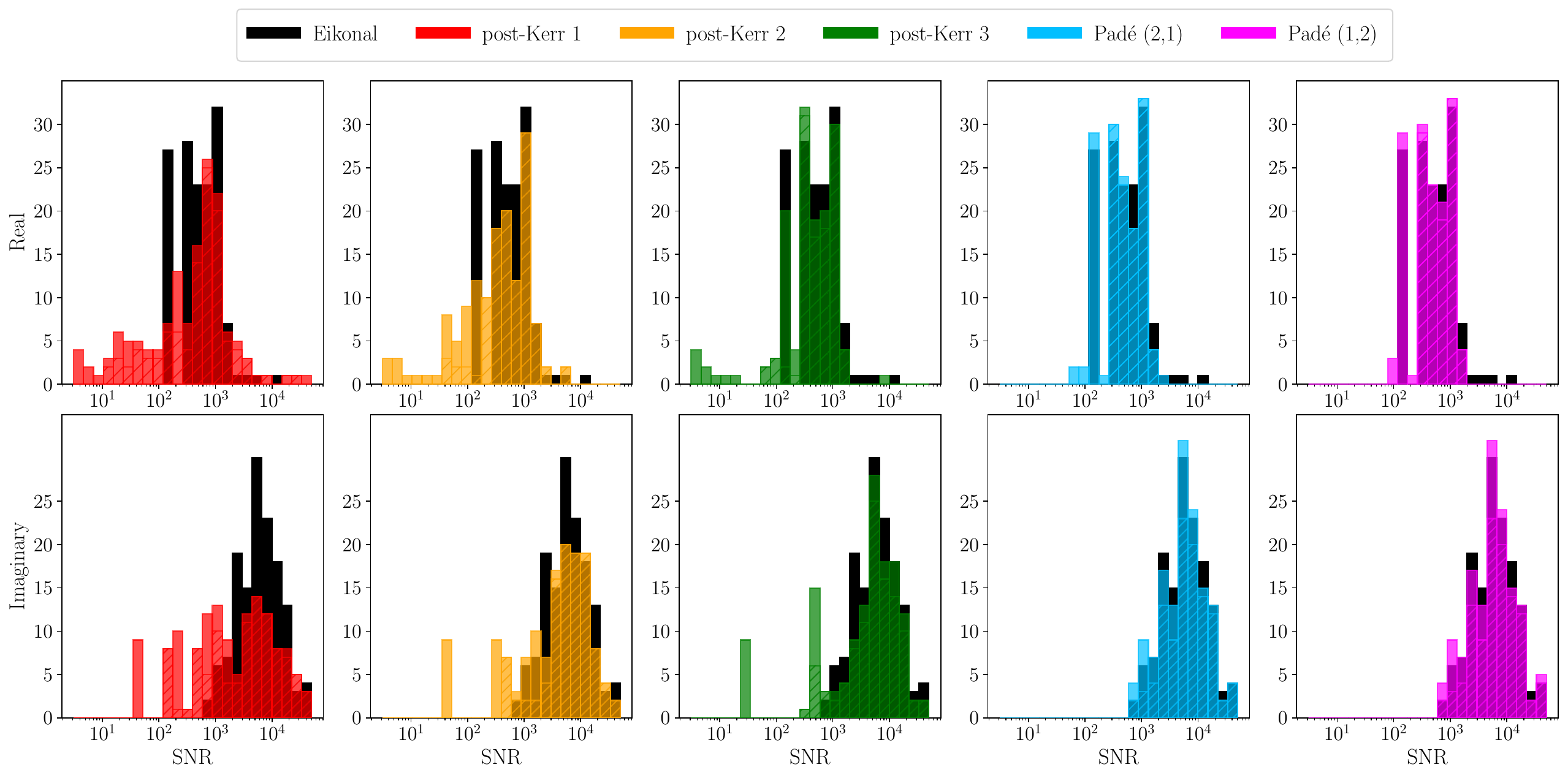}
\caption{Histograms of the SNR threshold obtained for prograde and retrograde modes for $a =  0.3\textrm{ and }0.7,\,\ell=m\in[2,10],\,\epsilon = 0.1,0.2,0.4 \textrm{ and } 1$. The eikonal predictions (black histogram) are used as a reference to assess the performance of the post-Kerr and Padé estimates. The hatched regions select only values of $\epsilon \leq 0.4$ where the post-Kerr approximation is more effective.}
\label{Fig: histograms}
\end{figure*}

Those SNR estimates are reported in Fig.~\ref{Fig: histograms} in a collection of histograms. 
The post-Kerr and Padé results are compared to the eikonal ones, used as a reference. The values of SNR corresponding to $\epsilon \neq 1$ are emphasized by hatching. 
This provides a visual understanding of the improvement of the post-Kerr approximation at higher orders, as well as its limitations for large values of $\epsilon$. 
In fact, the eikonal estimates appear well concentrated in a narrow range of large SNRs, while the post-Kerr estimates spread over a larger interval, especially at first order. 
One can also notice that the contribution at low SNR is mainly due to the $\epsilon=1$ component and becomes progressively smaller at larger post-Kerr orders.

Figure~\ref{Fig: histograms} also shows that the eikonal and post-Kerr approximations work better for the imaginary part, since the histograms are predominantly lying in areas of larger SNR. 
Comparing different post-Kerr orders, it is also clear that the overlap between the post-Kerr and the eikonal histograms increases with the order of approximation. 
The most significant improvement occurs between the first and second orders, while the third order only yields a small additional gain. 
On the other hand, the Padé approximants provide an extraordinary agreement with the eikonal results at both orders $(1,2)$ and $(2,1)$, even including the $\epsilon=1$ contribution. 
For existing data-analysis applications based on the eikonal approximation, e.g., Ref.~\cite{Dey:2022pmv}, currently accessible SNRs are well below our typical thresholds.

\section{Conclusions}\label{sec:conc}

As the precision of GW measurements and QNM extraction improves~\cite{Borhanian:2022czq,2025arXiv250908099T}, understanding how near-horizon modifications affect the QNM spectrum becomes a central question in theoretical astrophysics and BH spectroscopy. 
This question is particularly relevant because many alternatives to GR predict deviations confined to the near-horizon region, while leaving the asymptotic structure of the spacetime unchanged. 
As a first step in this direction, in this study, we have investigated a specific deformation of the Kerr BH, exploiting the eikonal/post-Kerr approximations and their connection to the photon sphere in order to study scalar QNMs in both frequency and time domains. 
This framework has allowed us to probe how BH deformations affect the fundamental QNMs and to explore the relation between the photon sphere and the event horizon. 

By performing $2+1$ numerical evolutions of the scalar perturbation equation, we have been able to perform scattering experiments between the modified Kerr BH and a Gaussian wave packet. 
The QNMs have then been extracted from the numerical waveforms using the Prony method and compared with the semi-analytic predictions given by the eikonal and post-Kerr approximations. 
We have shown the efficacy of the post-Kerr scheme at higher orders in the perturbation parameter and low spin, as well as its limitations for large spin or BH deformation. 

In particular, the first-order post-Kerr approximation is not very effective as an eikonal approximation even for relatively small deformations $\epsilon \leq 0.2$. 
Instead, higher orders consistently provide better results, especially at low spin. A further improvement can be attained via Padé approximants, as a way to obtain an efficient resummation of the third-order post-Kerr series even for large spin. 

From the statistical standpoint, the systematic error on the post-Kerr and eikonal estimates has been quantified by using the Prony values as a reference for the exact QNMs, while the statistical error has been obtained via MCMC sampling by injecting the Prony values into a single damped sinusoid. 
Combining the two types of error in the notion of bias ratio, we have been able to estimate the typical SNR up until the eikonal, post-Kerr, or Padé predictions can be used without introducing biases. 
This analysis has shown that the major improvement in the post-Kerr approximation occurs between first and second order, with a slight variation at third order. 
However, the Padé approximants consistently outperform any post-Kerr order and provide estimates that are very similar to the eikonal ones over the range of parameters considered in this paper. 

This work paves the way for many interesting developments. 
First of all, time domain simulations can be a useful tool to study higher overtones, which might be more sensitive to small deformations of the effective potential~\cite{Nollert:1996rf,Berti2022}, and modified theories of gravity~\cite{Konoplya:2022iyn,Nakashi:2025fbr}. 
Including matter to the 2+1 time-evolution code would also allow one to compare QNMs from linear perturbation theory with those extracted from numerical relativity simulations of BHs from merging and collapsing neutron stars~\cite{Steppohn:2025kbh}. 
It would also be interesting to apply the Padé resummation of the post-Kerr series to other BH metrics and eventually check its effectiveness in reproducing the eikonal results.

Additionally, this work could be generalized to gravitational perturbations and how they are affected by the near-horizon deformation. 
This requires first identifying a theory beyond GR, which admits the modified Kerr BH as a solution~\cite{Suvorov2020} and then studying the perturbation equation, which can be quite involved computationally. 
Gravitational perturbations would allow one to further test the QNM-photon sphere correspondence or test-field approximation for rotating BHs beyond GR~\cite{Li:2022pcy,Hussain:2022ins,Cano:2023tmv,Cano:2024wzo,Pani:2026yzi}, explore the reliability of these frameworks in the time-domain similar to Ref.~\cite{Thomopoulos:2025nuf}, and quantify QNM excitations and possible isospectrality loss~\cite{Silva:2024ffz,Silva:2026jih}. 
Furthermore, it would be intriguing to compare our methodology to the higher-order WKB (Wentzel–Kramers–Brillouin) method for rotating BHs~\cite{Seidel:1989bp,Kokkotas:1991vz,Tang:2025qaq}. 
Finally, it would also be interesting to further extend our analysis to the eikonal analysis of Refs.~\cite{Cano:2025mht,Cano:2025ejw}, which demonstrated that QNMs of rapidly rotating BHs in higher derivative gravity can be especially sensitive to small deviations from GR.

\acknowledgments
The authors thank Nicola Franchini for valuable feedback on the manuscript. 
The authors acknowledge support by the High Performance and Cloud Computing Group at the Zentrum für Datenverarbeitung of the University of Tübingen, the state of Baden-Württemberg through bwHPC and the German Research Foundation (DFG) through grant no INST 37/935-1 FUGG. 
C.\,D.\,S. is grateful to the University of T\"ubingen for hospitality during the realization of this work. 
C.\,D.\,S. and S.\,C. acknowledge the support of INFN, {\it sez. di Napoli}, {\it iniziative specifiche}  QGSKY and MoonLight2. 
S.\,H.\,V. acknowledges funding from the Deutsche Forschungsgemeinschaft (DFG): Project No. 386119226. 
V.\,D.\,F. and S.\,C. are grateful to the {\it Gruppo Nazionale di Fisica Matematica} (GNFM) of {\it Istituto Nazionale di Alta Matematica} (INDAM) for support. 
V.\,D.\,F. acknowledges the support of INFN, {\it sez. di Napoli}, {\it iniziativa specifica} TEONGRAV.

\bibliography{literature}

\appendix

\section{Numerical evolution and QNM extraction} \label{sec:appendix}

To implement the fourth-order Runge-Kutta algorithm, the KG equation is arranged in a system of first-order differential equations in the fields $\Psi$ and $\Pi=\partial_t\Psi$. 
The algorithm then evolves over time the quantity $u=\{\Psi_R,\Psi_I,\Pi_R,\Pi_I\}$, where the subscripts identify the real (R) and imaginary (I) parts of $\Psi$ and $\Pi$. 
The initial data have been chosen as a static Gaussian wave packet~\cite{Price:2004mm} modulated by a normalized associated Legendre polynomial in $\theta$, being the component of the spherical harmonics
    \begin{align}
        \Psi|_{t=0}&=P_\ell^m(\cos\theta)\,e^{-(r_*-r_*^0)^2/2}, \\
        \Pi|_{t=0}&=0\,.
    \end{align}
The Gaussian data are centered at $r_*^0=12M$, while the observer's position is at $r_*=30M$ and $\theta=\pi/2$. 
To obtain stable time evolutions, we have imposed ingoing boundary conditions near the horizon, $r_*\approx-100M$, and outgoing boundary conditions at infinity $r_*\approx 200M$. 
Moreover, we have set $M=1$ as the mass is a scaling factor. At the angular poles $\theta=0,\pi$ the conditions are: $\partial_\theta\Psi=0$ for $m=0$ and $\Psi=0$ for $m\neq 0$~\cite{Barack2007}. 

The numerical evolution has been performed on a two-dimensional grid $(r_*,\theta)$, choosing $N_r=7000$ points in the radial direction and $N_\theta=100$ for the angular part. 
We note that spherical harmonics are not eigenfunctions of the angular part of the $\Box$ operator on an axisymmetric spacetime, and this gives rise to a phenomenon called \emph{mode mixing}~\cite{Pazos-Avalos:2004uyd,Hod:1999rx}: an initial multipole $\ell$ will excite additional modes with the same value of $m$. 
However, we find that the most excited modes correspond to the prograde and retrograde modes associated with the initial spherical harmonics $P_\ell^m$, which are the ones we study in this paper, and are also not strongly affected by mode mixing. 

\begin{figure*}
    \centering
    \includegraphics[width=0.95\linewidth]{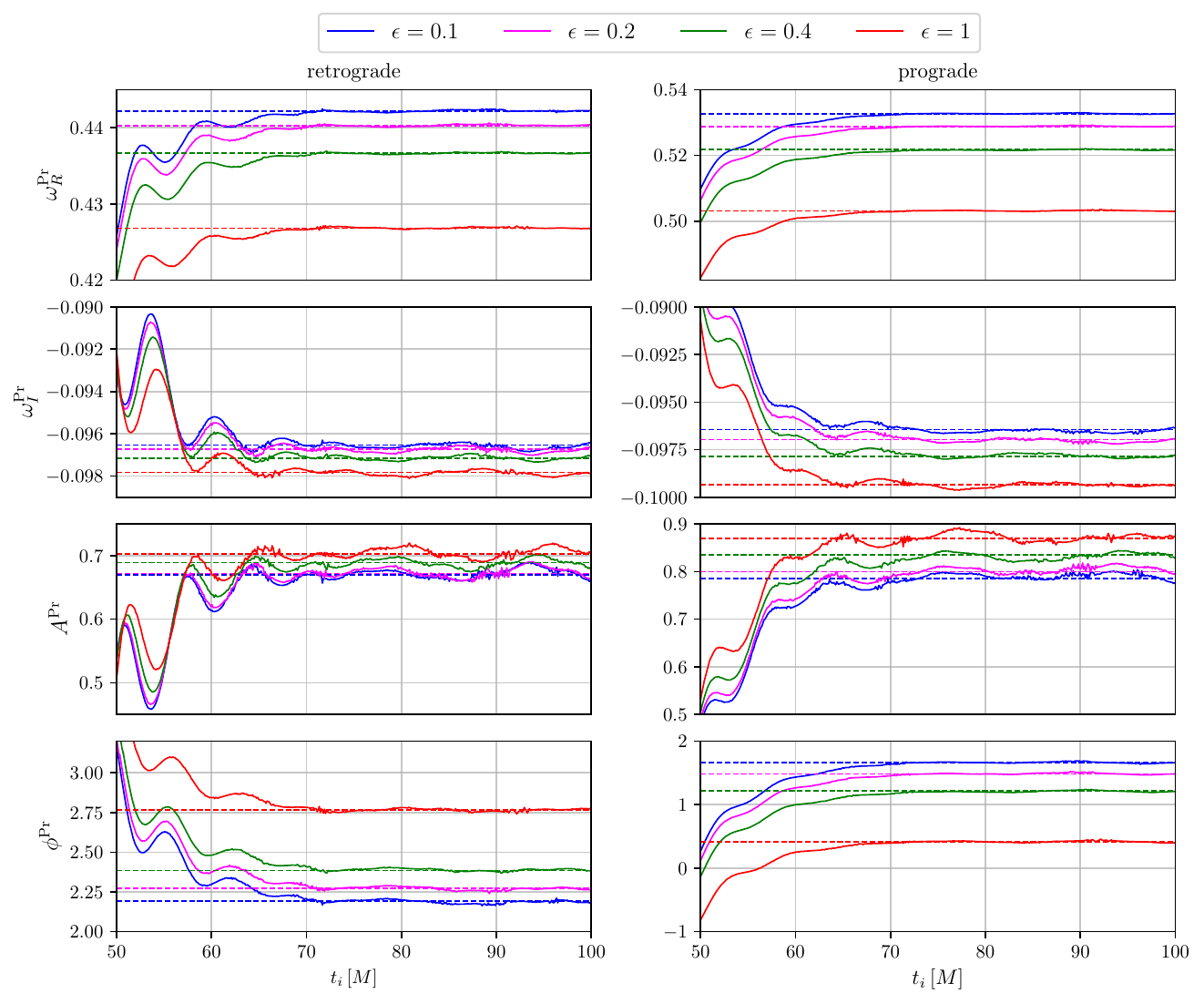}
    \caption{Prony estimates for prograde and retrograde modes as a function of the starting time $t_i$, example for $\ell=2$ and $a=0.3$. The dashed lines correspond to the average of the Prony estimates after convergence.
    The decaying oscillatory patterns as a function of the starting time of the fit agree qualitatively well with standard ringdown analysis papers.}  
    \label{fig:Prony_estimates}
\end{figure*}

\begin{figure*}
    \centering
    \includegraphics[width=0.95\linewidth]{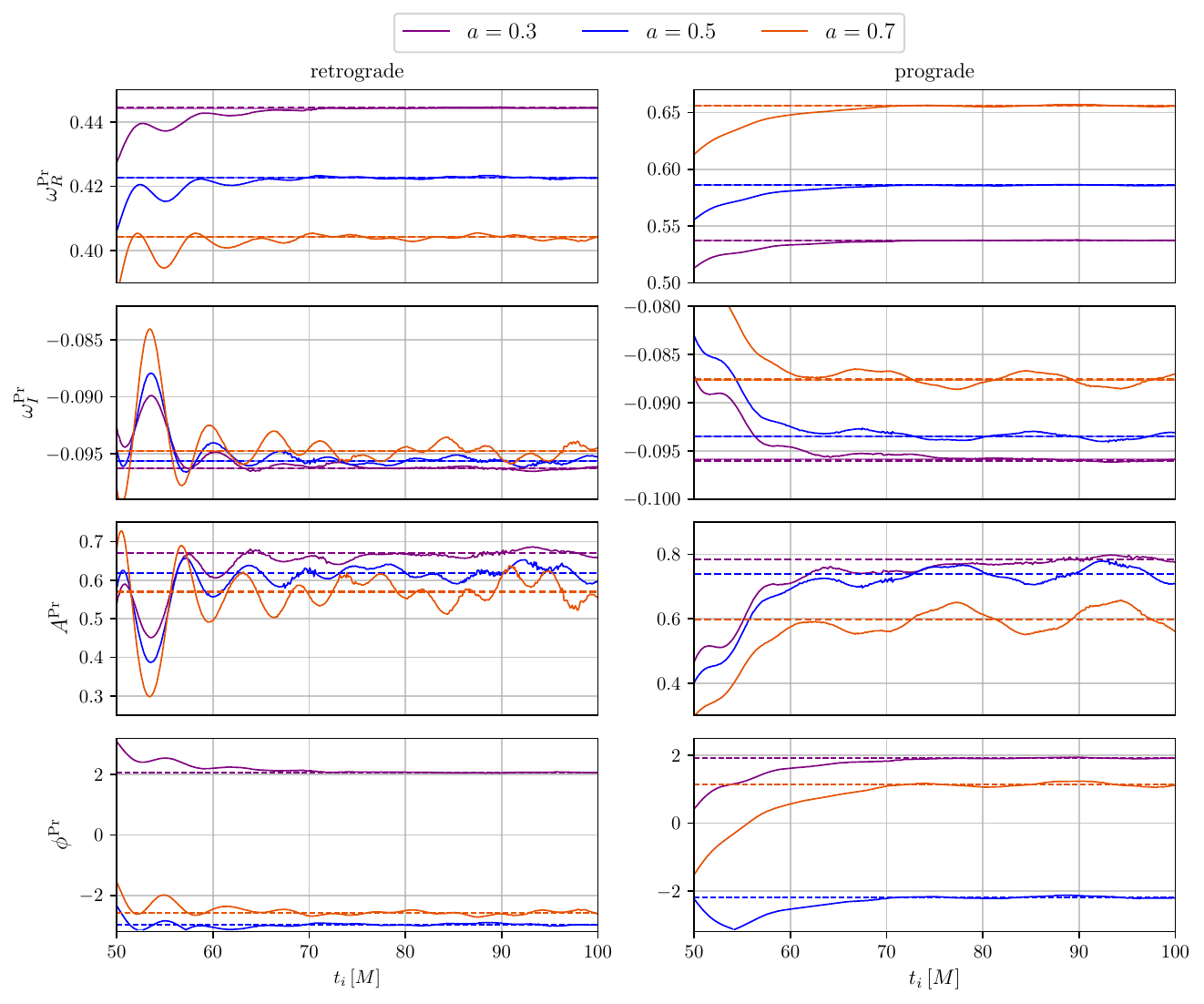}
    \caption{Prony estimates for prograde and retrograde modes of the Kerr BH as a function of the starting time $t_i$, example for $\ell=m=2$ and $a=0.3,0.5 \textrm{ and }0.7$. The dashed lines correspond to the average of the Prony estimates after convergence, while the solid lines in the first two rows are the exact scalar QNMs of the Kerr BH. The relative error of the QNM estimates compared to Leaver's method is $\leq 0.2\%$.}  
    \label{fig:Prony_estimates_Kerr}
\end{figure*}

The robustness of the Prony method can be assessed from Fig.~\ref{fig:Prony_estimates}, which reports the Prony estimates for the mode $(a=0.3,\,\ell=m=2)$ for selected values of $\epsilon$. It clearly shows that the frequencies, as well as amplitude and phase, converge to constant values at large starting times $t_i$ of the fit, as the oscillations in the Prony estimates become progressively smaller. For completeness, we have reported in Fig.~\ref{fig:Prony_estimates_Kerr} the convergence of the Prony estimates for the Kerr BH $(\epsilon=0)$ for different values of the BH spin. Here one can notice that the deviations from the average values tend to increase with the spin of the BH, especially in the amplitude,
as expected from the mode mixing effect. However, the relative error on the QNM estimates is always $\lesssim 0.2\%$. For the statistical analysis, we consider all QNMs, amplitude, and phase values obtained after convergence. 
The final Prony estimates are thus defined as the mean of these values, with statistical uncertainties given by the $90\%$ highest density interval. 
The errors obtained in this way vary around $10^{-3}$ and $10^{-5}$. 
Note that the oscillatory patterns are generally expected when a signal contains small but unmodelled contributions like overtones~\cite{Volkel:2025jdx}. 

\section{Data-analysis estimates} 
\label{sec:dataanalysis}

In the following, we outline how to estimate the statistical errors of QNM parameters and how to quantify the relevance of biases. 
We adopt standard techniques from GW data analysis. 

The waveform is a single damped sinusoid
\begin{equation}\label{model_waveform}
    h(t,\boldsymbol{\vartheta})=A \, e^{i\omega_R \,t+i\,\phi-\omega_I\,t}\,,
\end{equation}
where $\boldsymbol{\vartheta}=\{A,\phi,\omega_R,\omega_I\}$ corresponds to the model parameters of amplitude, phase, and frequencies. 
Given two waveforms $h_1(t)$ and $h_2(t)$, we define the inner product
    \begin{equation}
        \left< h_1 | h_2\right> = \frac{4}{S_h}\Re\left[\int_{t_i}^{t_f} h_1(t)\,h_2^*(t) \text{d} t\right]\,,
    \end{equation}
where the noise power-spectral density is assumed to be a constant $S_h$ over the integration domain $[t_i,t_f]$ (flat noise). 
In the absence of a specific detector in mind, we use this simple assumption as a default option. 

In terms of this inner product, the SNR $\rho$ of a numerical waveform $h(t)$ is defined as
    \begin{equation}
        \rho^2= \left< h | h\right>\,.
    \end{equation}
In the context of our theoretical study, the SNR of a given waveform can either be adjusted by changing the value of $S_h$ or the amplitude of the waveform. 

Next, we define the log-likelihood function
\begin{equation}
    \label{eq:likelihood}
    \ln p(d |\bm{\vartheta}) = -\frac{1}{2}\left< d-h(\bm{\vartheta}) | d-h(\bm{\vartheta}) \right>\,,
\end{equation}
which yields the probability of the data $d$ given the parameters $\bm{\vartheta}$. In our case, the data $d=h(t,\bm{\vartheta})$ is defined from Eq.~\eqref{model_waveform} by inserting the Prony estimates for $\boldsymbol{\vartheta}$.

For a given signal, the parameters $\boldsymbol{\vartheta}$ and their uncertainties $\Delta \omega$ are encoded in the posterior distribution, which can be inferred using Bayesian methods, like MCMC sampling and specifying a prior distribution. 
We adopt flat priors for all parameters and sample likelihood using the \texttt{BlackJAX} MCMC sampler~\cite{cabezas2024blackjax}. 
It is a high-performance library of sampling and inference algorithms based on \texttt{JAX}~\cite{jax2018github}, making it extremely fast and flexible. 

An alternative approach to approximate statistical errors of $\bm{\vartheta}$ evaluated in the known true parameters, is the Fisher-matrix framework, used to verify our MCMC findings. 
In this approach, the statistical uncertainties of the ringdown parameters $\bm{\vartheta}$ are approximated, in the high-SNR limit, by the covariance matrix $C \equiv \Gamma^{-1}\propto \rho^{-2}$. 

The diagonal elements of $C$, denoted by $\sigma_{ii}$, define the statistical uncertainties of the individual parameters as
$\Delta \vartheta_i \equiv \sqrt{\sigma_{ii}}$ are obtained by inverting the Fisher matrix $\Gamma$. 
The components of the Fisher matrix are given by
\begin{align}
\Gamma_{ij} = \left\langle \partial_i h \mid \partial_j h \right\rangle \,,
\end{align}
where the partial derivatives are taken with respect to the components of the set of parameters $\bm{\vartheta}$.

\section{Complementary results} \label{sec:complementary results}

\begin{figure*}
    \centering
    \includegraphics[width=\linewidth]{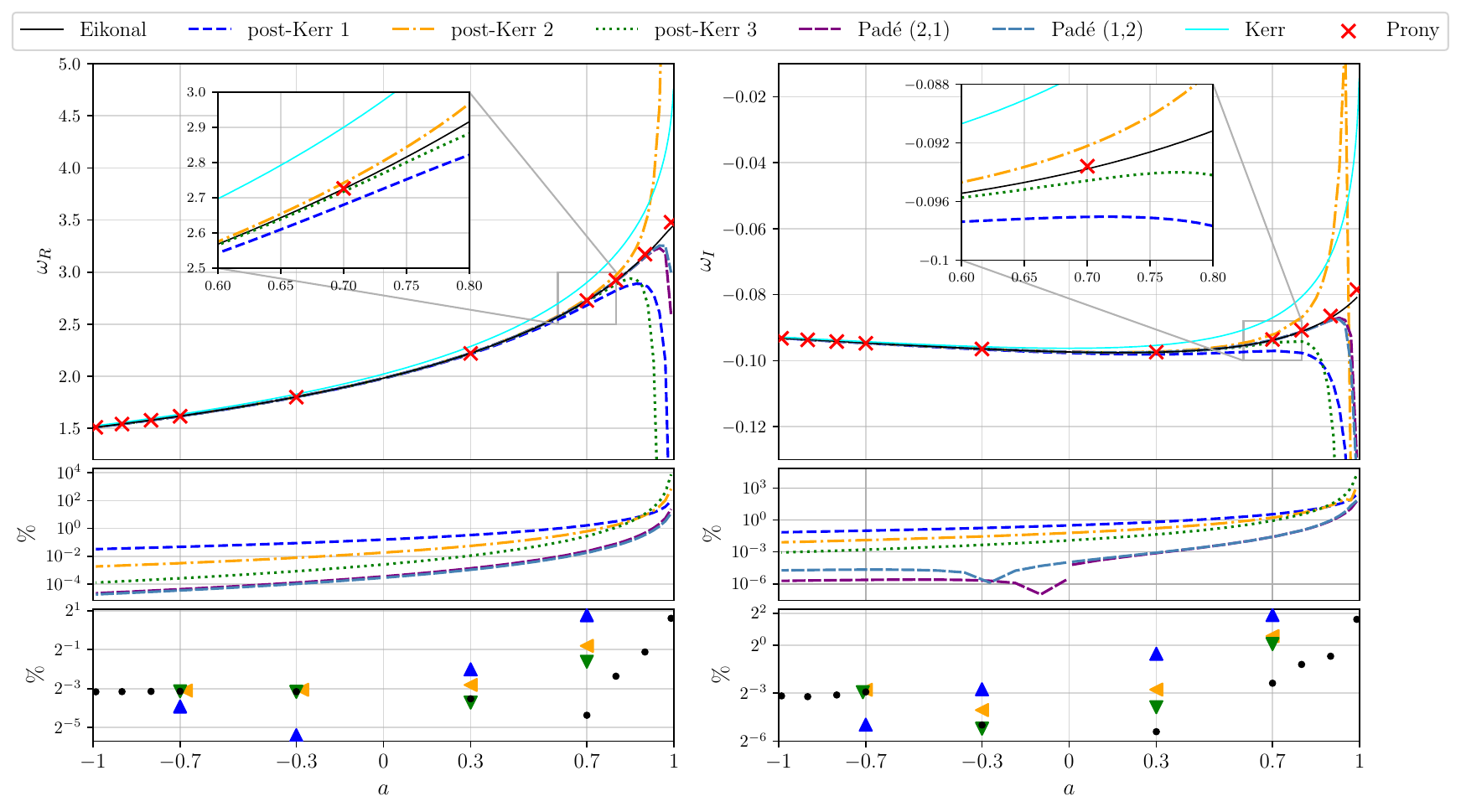}
    \caption{Same notation adopted in Fig.~\ref{fig:QNMs}, but using $\ell=10.$}
    \label{fig:QNMs_l10}
\end{figure*}

\begin{figure*}
\includegraphics[width=1.0\linewidth]{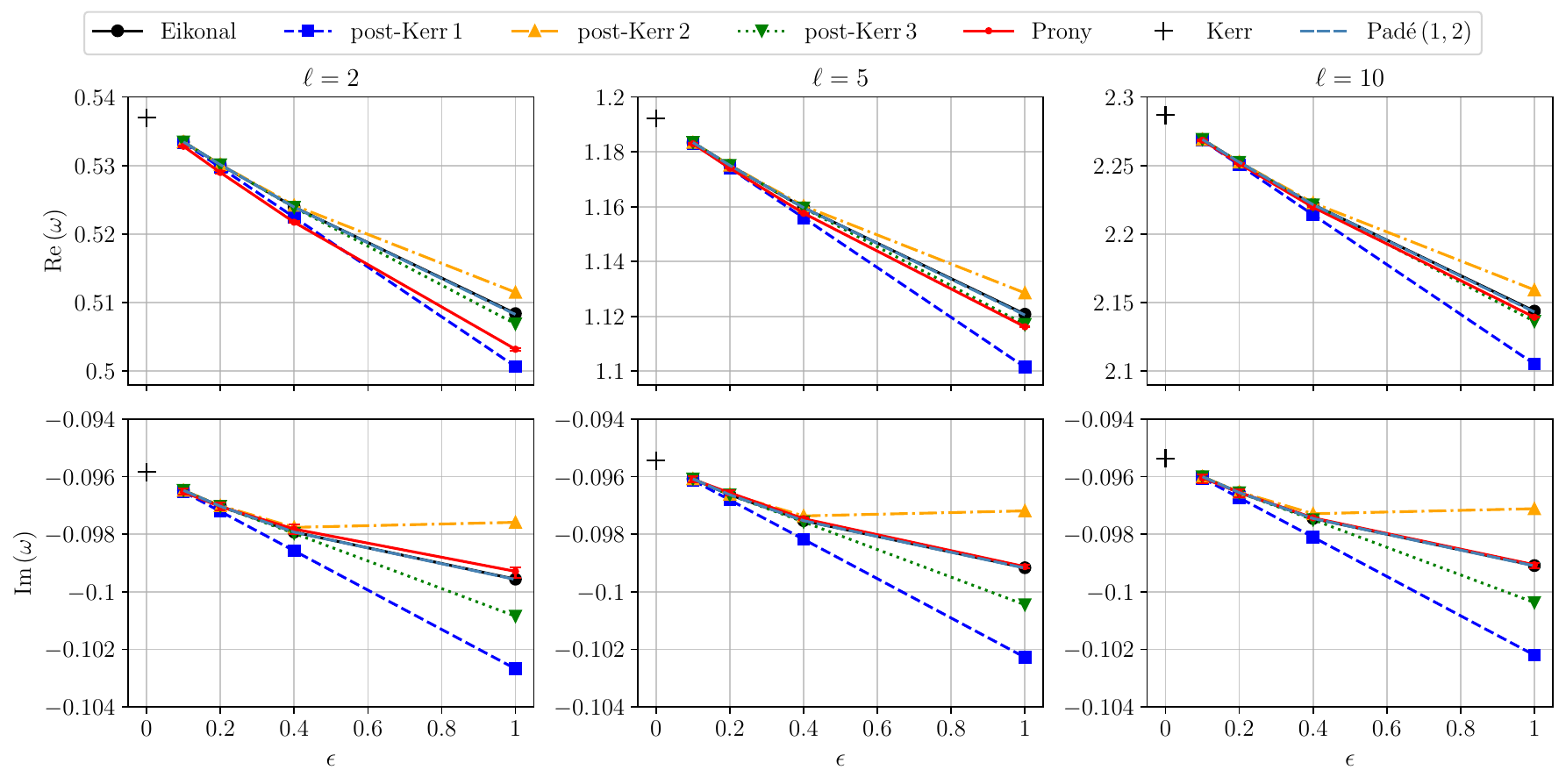}
\caption{Same notation adopted in Fig.~\ref{fig:pK_orders_07}, but using $a=0.3$.}
\label{fig:pK_orders_03}
\end{figure*}

This appendix contains two additional plots, which complement the results obtained in the main text. 
In particular, Fig.~\ref{fig:QNMs_l10} extends the results shown in Fig.~\ref{fig:QNMs} to the case $\ell=10$, where the eikonal approximation performs better. 
On the other hand, Fig.~\ref{fig:pK_orders_03} displays the Prony, post-Kerr, and eikonal prograde QNMs as a function of $\epsilon$ for spin $a=0.3$ and selected values of $\ell$. 

\end{document}